\documentclass[10pt,onecolumn,aps,prd,floats,showpacs,showkeys,floatfix,superscriptaddress,nofootinbib]{revtex4}
\usepackage{amsmath,amsfonts,dsfont}
\usepackage[dvips]{graphicx}
\usepackage{color}
\usepackage{epsfig}
\usepackage{hyperref}
\usepackage{mathrsfs}
\begin{document}
\title{Homotheties of a Class of Spherically Symmetric Space-Times \\ Admitting $G_3$ as Maximal Isometry Group}
\author{Daud Ahmad}
\thanks{daud.math@pu.edu.pk}
\affiliation {Department of Mathematics, University of the Punjab,
Lahore, Pakistan.}
\author{Kashif Habib}
\thanks{kashif.habib@bh.edu.pk}
\affiliation {Beacon House School System, Gujranwala, Pakistan.}
\begin{abstract}
The homotheties of spherically symmetric space–times admitting $G_4$, $G_6$ and $G_{10}$  as maximal isometry groups are already known, whereas for the space-times admitting $G_3$ as  isometry groups, the solution in the form of differential constraints on metric coefficients  requires  further classification.  For a class of spherically symmetric space-times admitting $G_3$ as maximal isometry groups without imposing any restriction on the stress-energy tensor, the metrics along with their corresponding homotheties are found. For the one case the metric is found along with its homothety vector that satisfies an additional constraint and is illustrated with the help of an example of a metric. For another case the metric and the corresponding homothety vector are found for a subclass of spherically symmetric space-times for which the differential constraint is reduced to separable form. Stress-energy tensor and related quantities of the metrics found are given in the relevant section.
\end{abstract}
\maketitle

\section{Introduction}
General Relativity (GR)  \cite{Schutz1985} formulates a  physical problem  in terms of differential equations as a geometric requirement  that a space-time may correspond to a Riemmannian manifold  as the interaction of matter and gravitation.  A relation between the geometry and the distribution of matter in the space-time is expressed by the  following  Einstein's  field equations $(EFEs)$~\eqref{efes},
\begin{equation}\label{efes}
    R_{ab}-\frac{1}{2} R g_{ab}+\Lambda g_{ab}\,=\,\kappa T_{ab}.
\end{equation}
GR  is expressed in terms of Pseudo-Riemannian Geometry. The torsion free space $(V_4)$ is represented by a Riemannian manifold $M$ of four dimensions having signature (+, -, -, -) with metric tensor $g_{ab}$ and symmetric stress  energy tensor $T_{ab}$ $(a, b \,=\, 0, 1, 2, 3)$. The curvature of the space-time is represented by the Riemann tensor $R_{abcd}$, where $R_{ab}\,=\,R_{acb}^{c}$ is the contraction of Riemann curvature tensor, $R \,=\, R_{a}^{a}$,  the Ricci scalar,  $\kappa\,=\,8\pi G/c^4$ and  $\Lambda$, the cosmological constant. The value of $\Lambda$ is observed to be negligible and usually taken to be zero. The $\Lambda$ term is only of significance at cosmological scale. In case of non-vanishing $\Lambda$, the $\Lambda$-term is usually treated as part of the stress-energy tensor $T_{ab}$.

EFEs \eqref{efes} break down into highly non-linear, second order coupled partial differential equations and  are difficult to handel in general unless certain symmetries are assumed by the space-times. Exact solutions ~\cite{Kramer1980} of EFEs \eqref{efes} may be found by requiring certain symmetry property of a space-time and they  have played a significant role in the discussion of physical problems e.g.  the Kerr  and Schwarzschild solutions  for the final collapsed state of massive bodies. The exact solutions mostly arose from highly idealized physical problems requiring  high symmetry as has been compiled by Kramer et al \cite{Kramer1980}, for example the well known spherically symmetric solutions of Schwarzschild, Reissner and Nordstrom, Tolman and Friedmann. The known exact solutions may be classified into (at least) four classes~\cite{Kramer1980}, namely the algebraic classification of conformal curvature, physical characterization of the energy momentum tensor,  existence and structure of preferred vector fields and  group of motions. The groups of symmetries  is used to construct more general cosmologies. One of these symmetries called homotheties  of a  space-time are more restrictive than   the isometries of the space-time. They are useful  to find the solutions of $EFEs$,  their properties and they can  model the  universe to find  new facts related to  cosmology and singularities  \cite{Eardley1974}. Classical Hydrodynamics also has  benefitted from the similarity solutions assuming the models for physical systems having no intrinsic scale of length, mass or time  \cite{Eardley1974}.  Cahill and Taub \cite{Cahill1971} analysed the homotheties of  the spherically symmetric distribution of self-gravitating perfect fluid  satisfying the homothety eqs.~\eqref{hvf}. Taub \cite{Taub1972} studied the homotheties of plane symmetric space-times underlining  the physical significance of homotheties in GR. Godfrey \cite{Godfrey1972} constructed all homothetic Wyel space-times.  Collinson and French \cite{Collinson1967}, Katzin, Lavine and Davis \cite{Katzin1969} and Collinson \cite{Collinson1970} studied more general geometric symmetries.  Roy Maartens et al~\cite{MaartensMaharajTupper1995} found the general solution of the conformal killing equations for the static spherically symmetric space-times and that for non-conformally-flat space-times, there are at the most two proper conformal motions including the regular conformally flat space-times with the conformal motion at the center. Farid, Qadir and Ziad \cite{TahaQadirZiad1995} classified static plane symmetric space–times according to their Ricci collineations (RCs) and  their relation with isometries of the space-times. Sharif and Sehar \cite{SharifSehar2005,SharifSehar2007} studied kinematic self-similar solutions of plane and cylindrically symmetric space-times for the perfect fluid and dust. Bokhari, Kashif and Qadir \cite{Kashif2005}, Saifullah and Yazdan \cite{Saifullah2009} studied conformal motions in the context of  plane symmetric static space-times.
Moopanar and Maharaj~\cite{Moopanar2010} investigated the conformal geometry of spherically symmetric distribution of mass, obtaining the general conformal Killing symmetry subject to a number of integrability conditions. They also found the rare space-times that admit the inheriting conformal symmetry given by their eq.~(11) for their sheering spherically symmetric space-times eq.~(1) through their general conformal symmetry eqs.~ $(5)$-$(6)$. Shabbir and Khan~ \cite{Shabbir2012} utilize the algebraic and direct integration techniques to find self-similar vector fields in static spherically symmetric space-times including the orthogonal, parallel and  non-parallel non-tilted proper self-similar vector fields for a special choice of the metric functions. Dorst~\cite{Dorst2016} explores the geometrical but computational way of working out conformal motions in 3D. Manjonjo et al~\cite{Manjonjo2017} studied the relationship between conformal symmetries and relativistic spheres in astrophysics and exploit  the non-vanishing components of the Weyl tensor to classify the conformal symmetries in static spherical space-times. Banerjee et al~\cite{Banerjee2017} explored the possibility of finding the static and spherically symmetric anisotropic compact stars in general relativity that admit conformal motions in the framework of $f(R)$ gravity theory.

As mentioned above, one of such restrictions could be to allow a space-time to admit certain symmetry properties. These symmetry properties lead a space-time to obey a certain Lie group or an isometry group. The isometry group $G_m$ of $(M,g)$ is the Lie group of smooth maps of $M$  into itself, leaving $g$ invariant. The subscript $m$ is equal to the number of generators or isometries of the group. It is the Lie algebra of continuously differentiable transformations $K={{K}^{a}}\left( {\partial }/{\partial {{x}^{a}}}\; \right)$, where $K^a \,=\, K^a(x^b)$ are the components of the vector field $K$, known as a Killing vector $(KV)$ field. A Killing vector field $K$ is a field along which the Lie derivative of the metric tensor $g$ is zero. i.e
\begin{equation}\label{kvf}
  \underset{K}{\mathscr{L}} g_{ab} \,=\,0,
\end{equation}
where ${\mathscr{L}}$ denotes the Lie derivative.  Besides isometries, there are symmetries called the self-similar solutions of  space-times which are more restrictive. These symmetry properties require a space-time to admit a Lie group, for example, the conformal motions, homothetic motions, Ricci collineations, curvature collineations, affine collineations etc. A homothety vector field  $H\,=\,H^a(\partial/\partial x^a)$ is a field along which the Lie derivative of a  metric tensor of a space-time remains invariant up to a scale, given by
\begin{equation}\label{hvf0}
  \underset{H}{\mathscr{L}} g_{ab} \,=\,  2\phi_{0} g_{ab},
\end{equation}
where $\phi_{0}$ is a scalar parameter, called the homothetic constant and $H$ the homothetic vector field.
Above eq.~\eqref{hvf0} may be rewritten in the  component form given below,
\begin{equation}\label{hvf}
   H^c \nabla _c g_{ab}+ g_{ac} \nabla_b H^c + g_{bc} \nabla_a H^c \,=\, 2\phi_{0} g_{ab}.
\end{equation}
 The corresponding homothety group is denoted by $H_{r}$, the subscript $r$ is the number of generators of the group. For $\phi_{0}\,=\,0$, the homotheties become motions but the converse may not be true.

It is well known  that for a Riemannian space $V_{n}$, the maximal group of motions is of the order less than or equal to $n\left( n\text{ }+\text{ }1 \right)/2$ . Fubini \cite{Fubini1903} has proved that a Riemannian manifold $V_{n}$ cannot admit a maximal group of the order $n\left( n+1 \right)/2-1$. Yegorov \cite{Yegorov1955} proved a result for Lorentzian manifolds, according to which the maximum group of mobility cannot be of the order$n\left( n+1 \right)/2-2$. It is well known \cite{Eardley1974} that for a Riemannian manifold with metric $g_{ab}$ and admitting $G_{m}$ as the maximal group of isometries, $H_{r}$ could be at the most of the order $r\,=\,m+1$. Thus for a $V_{n}$, $H_{r}$ could be at the most of the order $r\,=\,n(n+1)/2+1$. The  results of Fubini and Yegorov show that  a space–time $V_{n}$ cannot admit a homothety group $H_{r}$ with $r\,=\,n(n+1)/2-i$, where $i\,=\,0,1$. A detailed discussion of general relationship between isometries and homothetic motions can be seen in the  work \cite{HallandSteele1990,hall2004symmetries}. It is known that  for a $V_n$, there could be at the most $\left[ n\left( n+1 \right)/2 \right]+\text{ }1$ homothetic motions.  For the  spherically symmetric space-times
 \begin{equation}\label{eq:2.1.21}
  d{{s}^{2}}\,=\,{{e}^{\nu (t,r)}}d{{t}^{2}}-{{e}^{\lambda (t,r)}}d{{r}^{2}}-{{e}^{x(t,r)}}d{{\Omega }^{2}},
\end{equation}
 $d{{\Omega }^{2}}\,=\,d{{\vartheta }^{2}}+{{\sin }^{2}} \vartheta \, d{{\phi }^{2}}$, it is known \cite{AzadZiad1995} that the spherically symmetric space-times cannot admit a $G_5$ as the maximal group of motions. Qadir and Ziad   \cite{ZiadQadir1995} proved that the spherically symmetric space-times allow  isometry group $G_m$ of dimension $m\,=\,3,4,6,7,10$.  Therefore these space times could admit homothety groups $H_r$ of dimension $r\,=\,4,5,7,8,11$. To find  which space-time admits a non trivial homothety, the authors   \cite{DaudZiad1997}  solved the homothety equations for the spherically symmetric space-times for all the possible cases within the aforementioned  class of space-times. It came out that $r\neq 8$.  Thus,   for spherically symmetric space-times the possible maximal homothety groups $H_r$ could be of the order $r\,=\,4,5,7,11$. The solution of the homothetic eqs.~ \eqref{hvf}, as discussed in ref.  \cite{DaudZiad1997} results in the possible metrics along with the homothety groups $H_5$, $H_7$ and $H_{11}$.  However, for the space-times admitting $G_3$ as the maximal isometry group the solution  of the homothety eqs. ~\eqref{hvf} is provided   in the form of derivatives of unknown metric coefficients, which then requires a further classification for the case of homothety group $H_4$.

In order to find the space-times along with their respective homothety groups,  one needs to   solve the eqs.~ \eqref{hvf} for the spherically symmetric space-times \eqref{eq:2.1.21};  for details and background we refer the reader to  \cite{DaudZiad1997}. For $x\left( t,r \right)\, =\, \mu \left( t,r \right)+2 \ln r$, the solution  of the homothety eqs.~ \eqref{hvf} for the spherically symmetric space-times  \eqref{eq:2.1.21}  (dot and dash below denote derivatives $w.r.t.$  $t$ and $r$, respectively) is given by:
\begin{equation}
   H\, \,=\,\,   {{H}^{0}}\frac{\partial }{\partial t}+{{H}^{1}}\frac{\partial }{\partial r}+{{H}^{2}}\frac{\partial }{\partial \vartheta}+{{H}^{3}}\frac{\partial }{\partial \varphi},\label{eq:2.3.9}
\end{equation}
where
\begin{align}
  {{H}^{0}}\, =\, &-{{r}^{2}}{{e}^{\mu -\nu }}\left( \sin \vartheta \left( {{{\dot{g}}}_{1}}\sin \varphi -{{{\dot{g}}}_{2}}\cos \varphi  \right)+{{{\dot{g}}}_{3}}\cos \varphi  \right)+{{g}_{4}},\label{homothety0}\\
  {{H}^{1}}\, =\, &{{r}^{2}}{{e}^{\mu -\lambda }}\left( \sin \vartheta \left( {{{{g}'}}_{1}}\sin \varphi -{{{{g}'}}_{2}}\cos \varphi  \right)+{{{{g}'}}_{3}}\cos \vartheta  \right)+{{g}_{5}},\label{homothety1}\\
  {{H}^{2}}\, =\, &-\cos \vartheta \left( {{g}_{1}}\sin \varphi -{{g}_{2}}\cos \varphi  \right)+{{g}_{3}}\sin \vartheta +\left( {{c}_{1}}\sin \varphi -{{c}_{2}}\cos \varphi  \right),\label{homothety2}\\
  {{H}^{3}}\, =\, &-\cos ec\vartheta \left( {{g}_{1}}\cos \varphi +{{g}_{2}}\sin \varphi  \right)+\cot \vartheta \left( {{c}_{1}}\cos \varphi +{{c}_{2}}\sin \varphi  \right)+{{c}_{3}},\label{homothety3}
 \end{align}
 where ${{\text{c}}_{j}}\text{ }\left( j\,=\,1,2,3 \right)$  correspond to the generators of  $SO(3)$ and  ${{g}_{k}}$ for $k=1,2,3,4,5$ are functions of $t$ and $r$, and they  satisfy the following constraints:
 \begin{align}
 -\text{ }\dot{x}{{e}^{x-\nu }}\text{ }{{\dot{g}}_{j}}+{x}'{{e}^{x-\lambda }}\text{ }{{{g}'}_{j}}+\text{ 2}{{g}_{j}}\text{ }\,=\,&\text{ 0},\label{eq:2.3.1}\\
2{{\ddot{g}}_{j}}+( 2\dot{x}-\dot{\nu}){{\dot{g}}_{j}}-{\nu}'{{e}^{\nu-\lambda}}+\text{2}{{{g}'}_{j}}\text{}\,=\,&\text{0},\label{eq:2.3.2}\\
   2{{{\dot{g}}'}_{j}}+( {x}'-{\nu }' ){{\dot{g}}_{j}}+( \dot{x}-\dot{\lambda } ){{{g}'}_{j}}\text{ }\,=\,&\text{ 0},\label{eq:2.3.3}\\
  2{{{g}''}_{j}}+( 2{x}'-{\lambda }' ){{{g}'}_{j}}-{{e}^{\lambda -\nu }}{{\dot{g}}_{j}}\text{ }\,=\,&\text{ 0},\label{eq:2.3.4}\\
   \dot{x}{{g}_{4}}+{x}'{{g}_{5}}\text{ }\,=\,&\text{ 2}{{\varphi }_{0}},\label{eq:2.3.5}\\
  2{{\dot{g}}_{4}}+\dot{\nu }{{g}_{4}}+{\nu }'{{g}_{5}}\text{ }\,=\,&\text{ 2}{{\varphi }_{0}},\label{eq:2.3.6}\\
   {{e}^{\nu }}{{{g}'}_{4}}-{{e}^{\lambda }}{{\dot{g}}_{5}}\text{ }\,=\,&\text{ 0},\label{eq:2.3.7}\\
  2{{{g}'}_{5}}+{\lambda }'{{g}_{5}}+\dot{\lambda }{{g}_{4}}\text{ }\,=\,&\text{ 2}{{\varphi }_{0}}.\label{eq:2.3.8}
 \end{align}
A complete solution  of above eqs.~\eqref{eq:2.3.1}-\eqref{eq:2.3.8} provides all possible metrics with their homotheties admitting homothety groups $H_5$, $H_7$, $H_{11}$ except  for  homothety group $H_4$ for which the  space-times should admit additional differential constraints.  In \cite{DaudZiad1997}, it is found that for $r\,=\,5$ there are three space–times  (2.1)-(2.3) admitting ${{G}_{4}}\equiv SO(3)\otimes R$ where $R$ is time-like, space-like and null respectively, includes all static spherically symmetric space-times and the Bertotti–-Robinson metrics; for $r\,=\,7$ there are four space–times (Robertson–Walker ($RW$) space–times  (3.1) with   (3.5) and a $RW$-like space-time  (3.3) with  (3.9)) which admit ${{G}_{6}}\equiv SO(4),SO(3)\otimes {{R}^{3}},SO(1,3)$  as the maximal isometry groups; for $r\,=\,11$, the only space–time is  Minkowski space-time, for which  the maximal group of isometries is $SO(1,3)\otimes {{R}^{4}}$; for $r\,=\,4$, $H_4$ as the maximal group of homotheties the space-time admitting ${{G}_{3}}\equiv SO(3)$ as the maximal isometry group satisfy additional differential constraints ~(4.3)-(4.7), which require further classification according to different types of stress-energy tensor as  done by, e.g., Cahill and Taub~\cite{Cahill1971}.

In this paper,  we find  homotheties of a class of the  spherically symmetric space-times \eqref{eq:2.1.21} admitting ${{G}_{3}}\equiv SO(3)$ as the maximal isometry group  for  $x(t,r) \,=\, 2 \ln r$,  imposing no restriction on the stress-energy tensor. We accomplish this task in the  section~\ref{homothetiesH4}.  This gives rise to the two cases that either  $\dot{\lambda}\,=\,0$ or  $\dot{\lambda}\neq 0$. In the former case, the metric found is given by eq.~\eqref{eq:3.1.15} along with its homothety vector eq.~\eqref{eq:3.1.16}. In particular for $\nu (t,r)\,=\,\frac{t}{r},\hspace{0.25cm} h(t)\,=\,t$, the corresponding metric and the homothety vector are given by the eqs.~\eqref{eq:3.1.17} and \eqref{eq:3.1.18} respectively. In the latter  case, the metric and the homothety vector are given by the eqs.~\eqref{eq:3.1.36} and \eqref{eq:3.1.37}, for a subclass of spherically symmetric space-times~\eqref{eq:2.1.21} for  $\nu=\nu(r)$ for which one of the constraint equations is reduced to separable form. Furthermore,  we have included Ricci tensor components $R_{ab}$, Ricci scalar $R$ and the stress energy tensor $T_{ab}$ of the space-times ~\eqref{eq:3.1.15}, \eqref{eq:3.1.17}  and  \eqref{eq:3.1.36} in the relevant section. The results and remarks are presented in the final section~\ref{conclusion}.
\section{Spherically Symmetric Space-Times  Admitting $\mathbf{H_4}$ as the Homothety Group}\label{homothetiesH4}
For the space-times eq.~\eqref{eq:2.1.21} to have $SO(3)$ as the maximal isometry group, one must have $g_j (t, r) \,=\, 0$, for which eqs.~\eqref{eq:2.3.1}-\eqref{eq:2.3.4}  are satisfied, where $j=1,2,3$.  However, for a space-time to have homothetic motion, $g_4$ and $g_5$ satisfying the eqs.~\eqref{eq:2.3.5} - \eqref{eq:2.3.8} should include only one arbitrary constant, the $\phi _{0}$ corresponding to the scale parameter of the homothety. For the detail, we refer the reader to the ref.~\cite{DaudZiad1997}. Let us suppose that,
\begin{align}
  g_{4}\,=\,& \phi _{0}\, h(t,r),\label{eq:3.1.1a}\\
  g_{5}\,=\, & \phi _{0}\, g(t,r). \label{eq:3.1.1b}
\end{align}
For $g_j (t, r) \,=\, 0$ along with  $g_4$, and $g_5$ as given above, eqs.~\eqref{eq:2.3.5}--\eqref{eq:2.3.8} reduce to:
\begin{align}
    \dot{x}h+\dot{x}g\,=\,&2,\label{eq.3.1.2}\\
  2\dot{h}+\dot{v}h+{v}'g\text{ }\,=\,& 2,\\
  {{e}^{\nu }}{h}'-{{e}^{\lambda }}\dot{g}\,=\,& 0,\\
  2{g}'+{\lambda }'g+\dot{\lambda }h\,=\,& 2.\label{eq.3.1.5}
\end{align}
Corresponding homothety vector eq.~\eqref{eq:2.3.9}  in this case reduces to
\begin{equation}\label{eq:3.1.6}
 H\,=\,{{g}_{4}}\frac{\partial }{\partial t}+{{g}_{5}}\frac{\partial }{\partial r}+\left( {{c}_{1}}\sin \varphi -{{c}_{2}}\cos \varphi  \right)\frac{\partial }{\partial \vartheta }+\left( \cot \vartheta \left( {{c}_{1}}\cos \varphi +{{c}_{2}}\sin \varphi  \right)+{{c}_{3}} \right)\frac{\partial }{\partial \varphi }.
\end{equation}
As mentioned above, an attempt to solve  eqs.~\eqref{eq.3.1.2} to eq.~\eqref{eq.3.1.5}, with out imposing any restriction on stress-energy tensor $T_{ab}$ or line element, produces a solution in the form of differential constraints. However, for $x(t,r)=2 \ln r$  these differential constraints are reduced to meaningful expressions which then produce the metrics admitting the above mentioned homothety groups. For $ x \,=\, 2\ln r$,  the line element eq.~\eqref{eq:2.1.21}   comes out to be,
\begin{equation}\label{eq:3.1.7}
d{{s}^{2}}\,=\,{{e}^{\nu (t,r)}}d{{t}^{2}}-{{e}^{\lambda (t,r)}}d{{r}^{2}}-{{r}^{2}}d{{\Omega }^{2}},
\end{equation}
and    eqs.~\eqref{eq.3.1.2} to ~\eqref{eq.3.1.5}  yield,
\begin{align}
 g\,=\,&r,\label{eq:3.1.8}\\
  2\dot{h}+\dot{\nu }h+{\nu }'r\,=\,&2,\label{eq:3.1.9}\\
  {h}'\,=\,&0,\hspace{0.25cm} {{e}^{\nu }}\, \ne \, 0,\label{eq:3.1.10}\\
  {\lambda }'r+\dot{\lambda }h\,=\,&0.\label{eq:3.1.11}
\end{align}
From the eq.~\eqref{eq:3.1.11}, we find
\begin{equation}\label{eq:3.1.12}
  h\left( t,r \right)=- \frac{{\lambda }'r}{{\dot{\lambda }}}.
\end{equation}
For  $\dot{\lambda }\ne 0$, eq.~\eqref{eq:3.1.12} along with  eq.~\eqref{eq:3.1.10} gives us
\begin{equation}\label{eq:3.1.13}
  {{\left( \frac{{\lambda }'r}{{\dot{\lambda }}} \right)}^{\prime }}=0, \hspace{0.25cm} \dot{\lambda }\ne 0,\hspace{0.25cm} \lambda =\lambda \left(t,r \right).
\end{equation}
Here two cases arise, either $\dot{\lambda}=0$ or $\dot{\lambda}\neq 0$:
\begin{flushleft}
\textbf{Case 1}
\end{flushleft}
For  $\dot{\lambda}=0$, eq.~\eqref{eq:3.1.11} gives us ${\lambda }'=0$ as $r\neq 0$, which is possible only when   $\lambda =0$. Thus $\nu=\nu(t,r)$ and $\lambda=0$  reduce eq.~\eqref{eq:3.1.7} to the metric
\begin{equation}\label{eq:3.1.15}
d{{s}^{2}}={{e}^{\nu (t,r)}}d{{t}^{2}}-d{{r}^{2}}-{{r}^{2}}d{{\Omega }^{2}},
\end{equation}
that  satisfies the eqs.~\eqref{eq:3.1.8}, \eqref{eq:3.1.10} and \eqref{eq:3.1.11} with an additional constraint on $v\left( t,r \right)$ given by eq.~\eqref{eq:3.1.9}, where $h=h\left( t \right)$. The corresponding homothety vector $H$,  eq.~\eqref{eq:3.1.6} in this case reduces to
\begin{equation}\label{eq:3.1.16}
H={{\phi }_{0}}\left( h\left( t \right)\frac{\partial }{\partial t}+r\frac{\partial }{\partial r} \right)+\left( {{c}_{1}}\sin \varphi -{{c}_{2}}\cos \varphi  \right)\frac{\partial }{\partial \vartheta }+\left( \cot \vartheta \left( {{c}_{1}}\cos \varphi +{{c}_{2}}\sin \varphi  \right)+{{c}_{3}} \right)\frac{\partial }{\partial \varphi }.
\end{equation}
For the space-time eq.~\eqref{eq:3.1.15}, Ricci scalar and the independent non-zero components of the Ricci tensor are:
\begin{equation}\label{Rabeq:3.1.15}
  \begin{split}
    & {{R}_{00}}=\frac{1}{4r}\left( 4{\nu }'+r{{{{\nu }'}}^{2}}+2r{\nu }'' \right){{e}^{\nu (t,r)}},{{R}_{11}}=-\frac{1}{4}\left( {{{{\nu }'}}^{2}}+2{\nu }'' \right) \\
 & {{R}_{22}}=\frac{1}{2}r{\nu }',{{R}_{33}}=\frac{1}{2}r{\nu }'{{\sin }^{2}}\theta ,\, R=\frac{1}{2r}\left( 4{\nu }'+r{{{{\nu }'}}^{2}}+2r{\nu }'' \right),
  \end{split}
\end{equation}
the components of stress-energy tensor and the stress energy tensor $T$ for the above space-time eq.~\eqref{eq:3.1.15} are given below:
 \begin{equation}\label{Tabeq:3.1.15}
 \begin{split}
 & \kappa {{T}_{00}}=0,\kappa {{T}_{11}}=\frac{{{\nu }'}}{r},\kappa {{T}_{22}}=\frac{1}{4}r\left( {\nu }'\left( r{\nu }'+2 \right)+2r{\nu }'' \right), \\
 & \kappa {{T}_{33}}=\kappa {{T}_{22}}{{\sin }^{2}}\theta ,\kappa T=-\frac{{\nu }'\left( r{\nu }'+4 \right)}{2r}-{\nu }''.
 \end{split}
 \end{equation}
Thus the case for $\dot{\lambda}=0$ results in a  metric \eqref {eq:3.1.15}, which seems to be  uninteresting as the energy-density  of the corresponding self-gravitating system given by  eq.~\eqref{Tabeq:3.1.15} is  zero which is not possible for a physical system. This means that  every  temporal metric coefficient $\nu \left( t,r \right)$  for $\dot{\lambda}=0$  leads to  non-physical zones of the space-time.
For example, the space-time eq.~\eqref{eq:3.1.15}  for  $\nu (t,r)={t}/{r}$  and $h(t)=t$  reduces to
\begin{equation}\label{eq:3.1.17}
  d{{s}^{2}}={{e}^{{t}/{r}\;}}d{{t}^{2}}-d{{r}^{2}}-{{r}^{2}}d{{\Omega }^{2}}.
\end{equation}
Above space-time eq.~\eqref{eq:3.1.17} satisfies the additional constraint eq.~\eqref{eq:3.1.9}. In this case, the homothety vector $H$  eq.~\eqref{eq:3.1.16} of the above space-time eq.~\eqref{eq:3.1.17}     is given by
\begin{equation}\label{eq:3.1.18}
  H={{\phi }_{0}}\left( t\frac{\partial }{\partial t}+r\frac{\partial }{\partial r} \right)+\left( {{c}_{1}}\sin \varphi -{{c}_{2}}\cos \varphi  \right)\frac{\partial }{\partial \vartheta }+\left( \cot \vartheta \left( {{c}_{1}}\cos \varphi +{{c}_{2}}\sin \varphi  \right)+{{c}_{3}} \right)\frac{\partial }{\partial \varphi },
\end{equation}
clearly showing that the space-time eq.~\eqref{eq:3.1.17} admits four homotheties. For the space-time \eqref{eq:3.1.17}, Ricci scalar and the independent non-zero components of the Ricci tensor are:
\begin{equation}
\begin{split}
   & R_{00}=\frac{{{t}^{2}}}{4{{r}^{4}}}{{e}^{t/r}}, \,R_{11}=-\frac{t}{{{r}^{4}}}( r+\frac{t}{4}),\\& R_{22}=\frac{t}{2r},\, R_{33}={{R}_{22}}\,{{\sin }^{2}}\theta,  \, R = \frac{t^2}{2 r^4}.
\end{split}
\end{equation}
Using EFEs eqs.~\eqref{efes} (without cosmological constant), the components of stress-energy tensor and the stress energy tensor $T$ for the above space-time eq.~\eqref{eq:3.1.17} are given by:
\begin{equation}\label{eq1:section3.3}
\begin{split}
 & \kappa T_{00} = 0, \, \kappa T_{11} = -\frac{t}{r^3}, \,  \kappa T_{22} = \frac{t (2 r+t)}{4 r^2}, \\& \kappa T_{33} = \kappa T_{22} \sin ^2 \theta,   \kappa T = -\frac{t^2}{2 r^4}.
\end{split}
\end{equation}
As mentioned above, the case for $\dot{\lambda}=0$ results in a metric for which the energy-density of the spherically symmetric space-time is zero. In particular,  for  example for $\nu (t,r)={t}/{r}$,  the metric~\eqref {eq:3.1.15} reduces to the metric eq.~\eqref {eq:3.1.17} with stress-energy tensor components  eq.~\eqref{eq1:section3.3} corresponding to a self-gravitating-source with zero-energy density, a characteristic of the metric that  every choice of the temporal metric coefficient leads to some non-physical zones of the space-time.
\begin{flushleft}
\textbf{Case 2}
\end{flushleft}
Now for  $\dot{\lambda}\neq 0$,  eq.~\eqref{eq:3.1.13} suggests that
\begin{equation}\label{eq:3.1.19}
  \frac{{\lambda }'r}{{\dot{\lambda }}}=\alpha (t),
\end{equation}
where  $\alpha \left( t \right)$  depends on  $t$ only. Equation \eqref{eq:3.1.12} along with eq.~ \eqref{eq:3.1.19} yields
\begin{equation}\label{eq:3.1.20}
  h(t,r) = -\alpha (t).
\end{equation}
Substituting eq.~\eqref{eq:3.1.20} in eq.~\eqref{eq:3.1.9} yields
\begin{equation}\label{eq:3.1.21}
 -2\dot{\alpha }(t)+\dot{v}h+{v}'r=2.
\end{equation}
Note that eq.~ \eqref{eq:3.1.21} is separable for    $\dot{\nu}=0$. Thus,  we may rewrite above equation as,
\begin{equation}\label{eq:3.1.22}
\dot{\alpha }(t)=\frac{{v}'r-2\text{ }}{2},\text{ }\dot{\nu }=0.
\end{equation}
In the  above eq.~\eqref{eq:3.1.22},  the  $L.H.S.$  is function of time t only  whereas $R.H.S.$  is function of $r$ only, which is possible only when,
\begin{equation}\label{eq:3.1.23}
  \dot{\alpha }(t)=\frac{{\nu }'r-2}{2}={\alpha }_{1},
\end{equation}
where ${\alpha }_{1}$ is separation constant and the above eq.~\eqref{eq:3.1.23} can be solved by separating it  into two parts. We get
\begin{equation}\label{eq:3.1.26}
\alpha(t)={\alpha }_{1} t+{\alpha }_{2},
\end{equation}
where ${\alpha }_{2}$ is a constant of integration and,
\begin{equation}\label{eq:3.1.28}
  \nu (r)=2({\alpha }_{1}+1)\ln r.
\end{equation}
The eqs.~\eqref{eq:3.1.1a} and \eqref{eq:3.1.1b} along with eqs.~ \eqref{eq:3.1.8}, \eqref{eq:3.1.20} and  \eqref{eq:3.1.26}   reduce to
\begin{equation}\label{eq:3.1.27}
{{g}_{4}}=-{{\phi }_{0}}\left( {{\alpha }_{1}}t+{{\alpha }_{2}} \right),{{g}_{5}}={{\phi }_{0}}r.
\end{equation}
We find now  $\lambda(t,r)$. The eqs.~\eqref{eq:3.1.19} and ~\eqref{eq:3.1.26} together may be written as,
\begin{equation}\label{eq:3.1.29}
  {\lambda}' r = \dot{\lambda}({\alpha }_{1} t+{\alpha }_{2}).
\end{equation}
By the same argument used for eq.~\eqref{eq:3.1.23},  eq.~ \eqref{eq:3.1.29} is possible only when
\begin{equation}\label{eq:3.1.30}
  {\lambda}' r = \dot{\lambda}({\alpha }_{1} t+{\alpha }_{2})=m,
\end{equation}
where  $ m$ is a constant. Thus, from eq.~\eqref{eq:3.1.30} we get
\begin{equation}\label{eq:3.1.33}
  \lambda(r) = m\ln r  \hspace{0.25cm} \text{and} \hspace{0.25cm}  \lambda(t) = \frac{m}{{\alpha }_{1}} \ln \,({\alpha }_{1} t +{\alpha }_{2}).
\end{equation}
The  eq.~ \eqref{eq:3.1.33}  implies that,
\begin{equation}\label{eq:3.1.35}
    \lambda (t,r)\,=\, m\ln \text{ }r+\frac{m}{{{\alpha }_{1}}}\ln ({{\alpha }_{1}}t+{{\alpha }_{2}}), \, {{\alpha }_{1}}\, \neq \, 0.
\end{equation}
The eq.~\eqref{eq:3.1.7} along with  ~\eqref{eq:3.1.28} and \eqref{eq:3.1.35} is given by the following metric,
\begin{equation}\label{eq:3.1.36}
d{{s}^{2}} \,=\, {{r}^{2({{\alpha }_{1}}+1)}}d{{t}^{2}}-{{r}^{m}}{{({{\alpha }_{1}}t+{{\alpha }_{2}})}^{{m}/{{{\alpha }_{1}}}\;}}d{{r}^{2}}-{{r}^{2}}d{{\Omega }^{2}}.
\end{equation}
Thus the corresponding homothety vector H, eq.~\eqref{eq:3.1.6} is given by,
\begin{equation}\label{eq:3.1.37}
H=-{{\phi }_{0}}\left( {{\alpha }_{1}}t+{{\alpha }_{2}} \right)\frac{\partial }{\partial t}+\left( {{\phi }_{0}}r \right)\frac{\partial }{\partial r}+\left( {{c}_{1}}\sin \varphi -{{c}_{2}}\cos \varphi  \right)\frac{\partial }{\partial \vartheta }+\left( \cot \vartheta \left( {{c}_{1}}\cos \varphi +{{c}_{2}}\sin \varphi  \right)+{{c}_{3}} \right)\frac{\partial }{\partial \varphi}.
\end{equation}
Here the arbitrary constants are $\phi_0, c_1,c_2, c_3$ whereas ${\alpha }_{1}$ and ${\alpha }_{2}$ are constants on metric. So that the generators of homothety group $H_4$ are,
\begin{align}
   {{X}^{0}}=&-\left( {{\alpha }_{1}}t+{{\alpha }_{2}} \right)\frac{\partial }{\partial t}+r\frac{\partial }{\partial r},  \label{eq:3.1.38}\\
  {{X}^{1}}=&\sin \varphi \frac{\partial }{\partial \vartheta }+\cot \vartheta \cos \varphi \frac{\partial }{\partial \varphi },\label{eq:3.1.39}\\
    {{X}^{2}}=&-\cos \varphi \frac{\partial }{\partial \vartheta }+\cot \vartheta \sin \varphi \frac{\partial }{\partial \varphi },\label{eq:3.1.40}\\
   {{X}^{3}}=&\frac{\partial }{\partial \varphi },\label{eq:3.1.41}
\end{align}
where $[X^0,X^i]=0$ and $[X^i ,X^j] = X^k$ for $i.j,k = 0,1,2,3, i\neq j\neq k$. Showing that there are four homothety vectors given by eqs.~\eqref{eq:3.1.38}-\eqref{eq:3.1.41}. For the space-time eq.~\eqref{eq:3.1.36},  the independent non-zero components of the Ricci tensor and the Ricci scalar are:
\begin{equation}
\begin{split}
 & {{R}_{00}}={{\alpha }^{-2}}[-1+\frac{m{{\alpha }_{1}}}{2}+(1+{{\alpha }_{1}})(3{{r}^{-2({{\alpha }_{1}}-1)}}{{\alpha }^{\frac{2({{\alpha }_{1}}-1)}{{{\alpha }_{1}}}}}-{{r}^{2{{\alpha }_{1}}-1}}(1+{{\alpha }_{1}}){{\alpha }^{\frac{2{{\alpha }_{1}}-1}{{{\alpha }_{1}}}}}+ \\& \quad \qquad {{r}^{2{{\alpha }_{1}}-m}}(1-m+2{{\alpha }_{1}}){{\alpha }^{\frac{2{{\alpha }_{1}}-m}{{{\alpha }_{1}}}}})], \\&
 {{R}_{01}}=\frac{2}{r\alpha }, \\&
{{R}_{11}}={{r}^{-2}}[ 2+2(1+{{\alpha }_{1}})-{{(1+{{\alpha }_{1}})}^{2}}-{{r}^{-2({{\alpha }_{1}}-1)}}{\alpha^{\frac{-2({{\alpha }_{1}}-1)}{{{\alpha }_{1}}}}}+\frac{1}{2}m{{r}^{m-2{{\alpha }_{1}}}}(m-{{\alpha }_{1}}){\alpha^{\frac{m-2{{\alpha }_{1}}}{{{\alpha }_{1}}}}} ], \\&
{{R}_{22}}={{r}^{-3}}[ {{r}^{3}}-r{{\alpha }_{1}}{\alpha^{\frac{-2}{{{\alpha }_{1}}}}}+(m-1){{r}^{3-m}}{\alpha^{\frac{-m}{{{\alpha }_{1}}}}}+{\alpha^{\frac{-3}{{{\alpha }_{1}}}}}(1-3r{\alpha^{\frac{1}{{{\alpha }_{1}}}}}) ], \\&
{{R}_{33}}\,=\,\left( {{R}_{22}}+\frac{1-r{{\alpha }^{{{\alpha }_{1}}^{-1}}}}{{{r}^{4}}{{\alpha }^{4{{\alpha }_{1}}^{-1}}}} \right){{\sin }^{2}}\theta,\\&
R=  r^{-6}[-\frac{1}{2}{{m}^{2}}{{r}^{4-2{{\alpha }_{1}}}}{{\alpha }^{-2}}+{{\alpha }^{-\frac{m}{{{\alpha }_{1}}}}}{{r}^{4-m}}(3\alpha _{1}^{2}-3m-{{\alpha }_{1}}m+3{{\alpha }_{1}})+{{\alpha }_{1}}m{{r}^{4-2{{\alpha }_{1}}}}{{\alpha }^{-2}}+{{r}^{-2{{\alpha }_{1}}-m+6}} \\
 & \, \qquad {{\alpha }^{\frac{-2{{\alpha }_{1}}-m+2}{{{\alpha }_{1}}}}}-{{r}^{4-2{{\alpha }_{1}}}}{{\alpha }^{-2}}-2{{r}^{4}}-{{r}^{3}}{{\alpha }^{-\frac{1}{{{\alpha }_{1}}}}}(\alpha _{1}^{2}+2{{\alpha }_{1}}+1)+{{r}^{2}}{{\alpha }^{-\frac{2}{{{\alpha }_{1}}}}}(5{{\alpha }_{1}}+9)-r{{\alpha }^{-\frac{3}{{{\alpha }_{1}}}}}-{{\alpha }^{-\frac{4}{{{\alpha }_{1}}}}}].
\end{split}
\end{equation}
The  stress-energy tensor for the above space-time eq.~\eqref{eq:3.1.17} from EFEs eqs.~\eqref{efes} (without cosmological constant), comes out to be:
\begin{align}
  \kappa T_{00}  &=
       \begin{aligned}[t]\label{eq2:section3.3}
      & \frac{1}{4}[({{m}^{2}}-2){{\alpha }^{2}}-2{{\alpha }^{\frac{2-m}{{{\alpha }_{1}}}-2}}{{r}^{2-m}}+{{\alpha }^{-\frac{m+2}{{{\alpha }_{1}}}}}(2{{\alpha }^{\frac{2}{{{\alpha }_{1}}}}}{{r}^{2}}({{\alpha }_{1}}^{2}-m{{\alpha }_{1}}+3{{\alpha }_{1}}+m+2)-2{{\alpha }^{\frac{m}{{{\alpha }_{1}}}}}\\& {{r}^{m}}
({{\alpha }^{\frac{1}{{{\alpha }_{1}}}}}{{({{\alpha }_{1}}+1)}^{2}}r-{{\alpha }_{1}})){{r}^{2{{\alpha }_{1}}-m-2}}+2{{\alpha }^{-\frac{4}{{{\alpha }_{1}}}}}(2{{\alpha }^{\frac{4}{{{\alpha }_{1}}}}}{{r}^{4}}-3{{\alpha }^{\frac{2}{{{\alpha }_{1}}}}}{{r}^{2}}+{{\alpha }^{\frac{1}{{{\alpha }_{1}}}}}r+1){{r}^{2{{\alpha }_{1}}-4}}] ,
       \end{aligned}\\
\kappa T_{01} &=\label{kappaT01}
       \begin{aligned}[t]
\frac {2} {\alpha r},
\end{aligned}\\
\kappa T_{11}  & =
       \begin{aligned}[t]
&\frac{1}{4} [({{m}^{2}}-2){{\alpha }^{\frac{m-2{{\alpha }_{1}}}{{{\alpha }_{1}}}}}{{r}^{-2{{\alpha }_{1}}+m-2}}-2{{\alpha }^{\frac{m-4}{{{\alpha }_{1}}}}}{{r}^{m-6}}-2{{\alpha }^{\frac{m-3}{{{\alpha }_{1}}}}}{{r}^{m-5}}+18{{r}^{m-4}} {{\alpha }^{\frac{m-2}{{{\alpha }_{1}}}}} - \\&   2{{\alpha }^{\frac{m-1}{{{\alpha }_{1}}}}}{{r}^{m-3}}- 4{{\alpha }^{\frac{m}{{{\alpha }_{1}}}}}{{r}^{m-2}}+2\alpha _{1}^{2}{{r}^{-3}}(r-{{\alpha }^{\frac{m-1}{{{\alpha }_{1}}}}}{{r}^{m}})-6{{r}^{-2}} (m-2)  - 2{{\alpha }_{1}}\\&(2{{\alpha }^{\frac{m-1}{{{\alpha }_{1}}}}}{{r}^{m+1}}-5{{\alpha }^{\frac{m-2}{{{\alpha }_{1}}}}}{{r}^{m}}+m{{r}^{2}}-3{{r}^{2}}){{r}^{-4}}-2{{\alpha }^{\frac{2}{{{\alpha }_{1}}}-2}}{{r}^{-2{{\alpha }_{1}}}}],
       \end{aligned}\\
 \kappa T_{22} & =
 \begin{aligned}[t]
 &\frac{1}{2}[3{{r}^{-m}}{{\alpha }^{-\frac{m}{{{\alpha }_{1}}}}}+{{\alpha }^{-\frac{2}{{{\alpha }_{1}}}}}{{r}^{-2}}(3-2r{{\alpha }^{\frac{1}{{{\alpha }_{1}}}}})+m({{r}^{-2{{\alpha }_{1}}}}{{\alpha }^{-2}}-{{r}^{-m}}{{\alpha }^{-\frac{m}{{{\alpha }_{1}}}}})-{{2}^{-1}}(2+{{m}^{2}}) \\& {{r}^{-2{{\alpha }_{1}}}}{{\alpha }^{-2}}+{{r}^{2-m-2{{\alpha }_{1}}}}{{\alpha }^{-2+\frac{2-m}{{{\alpha }_{1}}}}}-{{\alpha }^{-\frac{4}{{{\alpha }_{1}}}}}(-1+r{{\alpha }^{\frac{1}{{{\alpha }_{1}}}}}+3{{r}^{2}}{{\alpha }^{\frac{2}{{{\alpha }_{1}}}}}-{{r}^{3}}{{\alpha }^{\frac{3}{{{\alpha }_{1}}}}}){{r}^{-4}}- \\& (2+m){{r}^{-m}}{{\alpha }^{-\frac{m}{{{\alpha }_{1}}}}}+{{\alpha }_{1}}-({{\alpha }^{-\frac{1}{{{\alpha }_{1}}}}}{{r}^{-1}}-3{{r}^{-m}}{{\alpha }^{-\frac{m}{{{\alpha }_{1}}}}})\alpha _{1}^{2}],
         \end{aligned}\\
 \kappa T_{33} &=
 \begin{aligned}[t]\label{eq3:section3.3}
  &\left[ \kappa {{T}_{22}}+{{\alpha }^{-\frac{4}{{{\alpha }_{1}}}}}{{r}^{-4}}\left( 1-r{{\alpha }^{\frac{1}{{{\alpha }_{1}}}}} \right) \right]{{\sin }^{2}}\theta,
       \end{aligned}\\
   \kappa T &=
\begin{aligned}[t]\label{eq4:section3.3}
 & {{r}^{-6}}[\frac{1}{2}{{m}^{2}}{{r}^{4-2{{\alpha }_{1}}}}{{\alpha }^{-2}}-{{\alpha }^{-\frac{m}{{{\alpha }_{1}}}}}{{r}^{4-m}}(3\alpha _{1}^{2}-3m-{{\alpha }_{1}}m+3{{\alpha }_{1}})-{{\alpha }_{1}}m{{r}^{4-2{{\alpha }_{1}}}}{{\alpha }^{-2}}-\\
 & {{r}^{-2{{\alpha }_{1}}-m+6}}   {{\alpha }^{\frac{-2{{\alpha }_{1}}-m+2}{{{\alpha }_{1}}}}}+{{r}^{4-2{{\alpha }_{1}}}}{{\alpha }^{-2}}+2{{r}^{4}}+{{r}^{3}}{{\alpha }^{-\frac{1}{{{\alpha }_{1}}}}}(\alpha _{1}^{2}+2{{\alpha }_{1}}+1)-{{r}^{2}}{{\alpha }^{-\frac{2}{{{\alpha }_{1}}}}}\\
 &(5{{\alpha }_{1}}+9)+ r{{\alpha }^{-\frac{3}{{{\alpha }_{1}}}}}+{{\alpha }^{-\frac{4}{{{\alpha }_{1}}}}}].
\end{aligned}
\end{align}
The expression for the stress-energy tensor shows that the symmetries assumed correspond to more  complicated dynamics in the case of $H_{4}$ for $\dot{\lambda}\neq 0$ (eq.~\eqref{eq:3.1.13}). The  metric in this case  eq.~\eqref{eq:3.1.36} can be further analysed for different types of stress-energy tensor for which the corresponding homothety vector is given by eq.~\eqref{eq:3.1.37} and the stress-energy tensor and its components  by the eqs.~\eqref {eq2:section3.3} to \eqref {eq3:section3.3} and \eqref {eq4:section3.3}. For example, the presence of the term $T_{01}$ suggests the heat dissipation effects in the evolution of self-similar space-time. The role of heat dissipation has utmost relevance in the study of one of the interesting phenomena of the stellar structures, $i.e.$,  the behavior of spherically symmetric gravitational collapse of stellar masses. This process is highly dissipative in nature \cite{Bhatti2017}.  In order to see the role of radiations in the gravitational implosion, one must has non-zero values of $T_{01}$. One can see from eq.~\eqref{kappaT01} that  the quantity $\alpha(t)={\alpha }_{1} t+{\alpha }_{2}$ given by eq.~\eqref{eq:3.1.26} is controlling the effects of heat dissipation in the evolution of self-gravitating system. It can also be seen that  $\alpha(t)={\alpha }_{1} t+{\alpha }_{2}$ (eq.~\eqref{eq:3.1.26}) depends further on two  integration constants, i.e., $\alpha_1$ and $\alpha_2$. The presence of  these two constants $\alpha_1$ and $\alpha_2$ in eq.~\eqref{kappaT01} indicates the  dissipation effects in our analysis. It is interesting to mention that one cannot see the dynamics of non-radiating system from our results, as $\alpha_1$ and $\alpha_2$ cannot be zero simultaneously (as such a choice will assign undefined values to $T_{01}$). However, for  $\alpha_2=0$, $\kappa T_{00}$ reduces to  ${2}/{rt{{\alpha }_{1}}}$, thereby indicating the dependence of radial metric coefficient  on both $t$ and $r$.
\section{Conclusions} \label{conclusion}
We have found  the homotheties and corresponding metrics for a class of  spherically symmetric space-times  ~\eqref{eq:2.1.21} to admit $G_3$ as the maximal group of motions for   $x(t,r) = 2 \ln r$. The motivation behind is the  classification of spherically symmetric space-times according to their homotheties without imposing any restriction on the stress-energy tensor. The homotheties and the corresponding metrics are already known for the space-times admitting $G_4$, $G_6$ and $G_{10}$  as maximal isometry groups whereas for the space-times admitting $G_3$ as the isometry group the solution is known  in the form of differential constraints which needs further consideration. We found the homotheties and the corresponding metrics admitting  $G_3$ as the maximal group of isometries for a class of  spherically symmetric space-times  ~\eqref{eq:2.1.21}, as mentioned above. For a subclass of spherically symmetric space-times, for which $\dot{\lambda}=0$, the  metric is given by eq.~\eqref{eq:3.1.15} and the corresponding homothety vector is given by eq.~\eqref{eq:3.1.16} subject to the additional constraint eq.~\eqref{eq:3.1.9} in terms of derivatives of the metric coefficients. In particular for $\nu (t,r)\,=\,\frac{t}{r},\hspace{0.25cm} h(t)\,=\,t$, the  metric satisfying the additional constraint eq.~\eqref{eq:3.1.9} and the corresponding homothety vector are given by the eqs.~\eqref{eq:3.1.17} and \eqref{eq:3.1.18} respectively. Whereas, for  $\dot{\lambda}\neq0, \dot{\nu}=0$ the metric eq.~\eqref{eq:2.1.21} reduces to the metric eq.~\eqref{eq:3.1.36} and the corresponding homothety vector is eq.~\eqref{eq:3.1.37}. Stress-energy tensor for the above space-times are given by the eqs.~\eqref{eq1:section3.3} and ~\eqref{eq2:section3.3}- \eqref{eq4:section3.3}.  It might be interesting to see these results according to different types of stress-energy tensor.
\section{Conflicts of Interest}
The authors declare that there is no conflict of interest regarding the publication of this paper.
\bibliographystyle{unsrt} 
\bibliography{xbib}

\end{document}